\documentclass[vecphys]{svmult}

\usepackage{makeidx}         
\usepackage{graphicx}        
\usepackage{multicol}        
\usepackage{cite}            
\usepackage[bottom]{footmisc}
\usepackage{amsmath}
\usepackage{bm}
\usepackage{amstext}
\usepackage{amsxtra}
\usepackage{amssymb}

\begin{document}

\newcommand{\To}{T_c^0}
\newcommand{\kB}{k_{\rm B}}
\newcommand{\dT}{\Delta T_c}
\newcommand{\lo}{\lambda_0}
\newcommand{\cs}{$\clubsuit$}
\newcommand{\thold}{t_{\rm hold}}
\newcommand{\Nmf}{N_c^{\rm MF}}
\newcommand{\Tmf}{T_c^{\rm MF}}
\newcommand{\el}{\gamma_{\rm el}}

\newcommand{\Nc}{N_{\rm c}}
\newcommand{\Ncid}{N_{\rm c}^{0}}
\newcommand{\Nt}{N}
\newcommand{\aho}{a_{\rm ho}}
\newcommand{\Tcid}{T_{\rm c}^{0)}}

\newcommand{\mumf}{\mu_c^{\rm MF}}
\newcommand{\nmf}{n_c^{\rm MF}}
\newcommand{\gammael}{\gamma_{\rm el}}

\title*{Effects of interactions on Bose-Einstein condensation of an atomic gas}
\titlerunning{Bose-Einstein condensation of interacting atoms}
\author{Robert P. Smith and Zoran Hadzibabic}
\institute{Cavendish Laboratory, University of Cambridge, J. J. Thomson Avenue, Cambridge, CB3 0HE, United Kingdom
}

\setcounter{tocdepth}{3}

\maketitle

\begin{abstract}
The phase transition to a Bose-Einstein condensate is unusual in that it is not necessarily driven by inter-particle interactions but can occur in an ideal gas as a result of a purely statistical saturation of excited states.  However, interactions are necessary for any system to reach thermal equilibrium and so are required for condensation to occur in finite time.  In this Chapter we review the role of interactions in Bose-Einstein condensation, covering both theory and experiment.  We focus on measurements performed on harmonically trapped ultracold atomic gases, but also discuss how these results relate to the uniform-system case, which is more theoretically studied and also more relevant for other experimental systems.

We first consider interaction strengths for which the system can be considered sufficiently close to equilibrium to measure thermodynamic behaviour. In particular we discuss the effects of interactions both on the mechanism of condensation (namely the saturation of the excited states) and on the critical temperature at which condensation occurs.
We then discuss in more detail the conditions for the equilibrium thermodynamic measurements to be possible, and the non-equilibrium phenomena that occur when these conditions are controllably violated by tuning the strength of interactions in the gas.

\end{abstract}

\contentsline {title}{Contents}{1}
\contentsline {section}{\numberline {1}Introduction}{2}
\contentsline {subsection}{\numberline {1.1}Noninteracting Bosons}{2}
\contentsline {subsection}{\numberline {1.2}Interacting Bosons}{4}
\contentsline {subsection}{\numberline {1.3}Chapter outline}{5}
\contentsline {section}{\numberline {2}Precision measurements on a Bose gas with tuneable interactions }{6}
\contentsline {section}{\numberline {3}Non-saturation of the excited states}{6}
\contentsline {section}{\numberline {4}Interaction shift of the transition temperature}{10}
\contentsline {subsection}{\numberline {4.1}Measurements on a harmonically trapped Bose gas}{12}
\contentsline {subsection}{\numberline {4.2}Connection with a uniform Bose gas}{13}
\contentsline {section}{\numberline {5}Equilibrium criteria and non-equilibrium effects}{16}
\contentsline {section}{References}{18}

\section{Introduction}
\label{sec:intro}

Virtually all thermodynamic phase transitions are driven by interactions between particles, which promote symmetry breaking into an ordered state. The phase transition comes about as a result of the competition between the energy, which favours the ordered state, and the entropy, which favours the disordered state.
In contrast, Bose-Einstein condensation (BEC) is a purely statistical phase transition, which at least in principle should not rely on interactions. The transition is instead a direct consequence of the finite-temperature saturation of the number of particles in the excited states of the system \cite{Einstein:1925a, huan87,Pethick:2002,Pitaevskii:2003}.
While this statistical argument does not explicitly invoke interactions between the particles, it does assume that the gas is in thermal equilibrium, which is impossible to attain in a completely noninteracting system\footnote{In the recently observed Bose-Einstein condensation of a photon gas \cite{Klaers:2010}, there is no direct interaction between the light particles. However the interaction with the material environment, which ensures thermalisation, leads to a second-order interaction between the photons.}. This makes it challenging to experimentally observe ideal-gas behaviour and disentangle the role of interactions on the thermodynamics and dynamics of condensation.

In this Chapter we review our recent experiments on this topic \cite{Tammuz:2011,Smith:2011,Smith:2011b}, performed with an ultracold Bose gas of $^{39}$K atoms with tuneable interactions. We were able to identify the interaction regime in which the gas may be considered to be in thermal equilibrium and also to extrapolate our results to the noninteracting limit where direct equilibrium measurements are not possible. This allowed us to verify the statistical-saturation BEC mechanism in the noninteracting limit, and to accurately determine the deviations from ideal-gas behaviour due to interactions; these are seen both in the non-saturation of the excited states and in the shift of the critical point. Before presenting the experimental results we briefly review some background theory that will be useful for our discussion.

\subsection{Noninteracting Bosons}

We start by considering an ideal, noninteracting Bose gas. We first derive the key results for a uniform system, which we then apply to the trapped gas using the local density approximation (LDA).
This ``local" approach will be useful later when we consider the effects of interactions, and in particular for the comparison of a uniform Bose gas with one that is harmonically trapped.

The equilibrium momentum distribution of noninteracting bosons with mass $m$ at a temperature $T$ is given by the Bose distribution function
\begin{equation}\label{eq:bosep}
    f_p=\frac{1}{\mathrm{e}^{(p^2/2m-\mu)/\kB T}-1} \; ,
\end{equation}
where $p$ is the momentum and $\mu \leq 0$ the chemical potential.
The total particle density $n$ can be found by integrating over all momentum states:
\begin{equation}\label{eq:densityint}
    n=\int\frac{\mathrm{d}\mathbf{p}}{(2\pi \hbar)^3}\frac{1}{\mathrm{e}^{(p^2/2m-\mu)/\kB T}-1}=\frac{g_{3/2}(\mathrm{e}^{\mu/\kB T})}{\lambda^3} \; ,
\end{equation}
where $g_{3/2}(x)=\sum_{k=1}^{\infty}x^{k}/k^{3/2}$ is a polylogarithm function and  $\lambda \; = \; [2\pi\hbar^2/(m \kB T)]^{1/2}$ is the thermal wavelength. We can re-express this result in terms of the phase space density $D$ as
\begin{equation}\label{eq:psd}
    D\equiv n\lambda^3=g_{3/2}(\mathrm{e}^{\mu/\kB T}) \; .
\end{equation}
Eq.~(\ref{eq:psd}) shows that there is a maximum value that $D$ can take. This critical value is reached when     $\mu=0$ and is given by $D_c=g_{3/2}(1)=\zeta(3/2)\approx2.612$ (where $\zeta$ is the Riemann function). At a given temperature this corresponds to a maximum density.
If this density is reached all the excited states saturate and any additional particles must accumulate in the ground state, forming a Bose-Einstein condensate\footnote{The singular ground-state contribution to the total density is implicitly excluded from the integral in Eq.~(\ref{eq:densityint}).  As $\mu$ approaches zero from below the ground state occupation can become arbitrarily large, as can be seen by inspecting Eq.~(\ref{eq:bosep}).}.
At a given density $n$ the BEC transition temperature is given by
\begin{equation}\label{eq:uniformTc}
   \kB T_c^0=\frac{2\pi \hbar^2}{m}\left(\frac{n}{\zeta(3/2)}\right)^{2/3} \; ,
\end{equation}
where the superscript $^0$ refers to the fact this is an ideal gas result.

For a gas in a potential $V(\mathbf{r})$ we may apply LDA to Eq.~(\ref{eq:densityint}). This amounts to having a local chemical potential
\begin{equation}\label{eq:mulocal}
\mu(\mathbf{r})=\mu-V(\mathbf{r}) \; .
\end{equation}
Specifically for a harmonic trap, $V(\mathbf{r}) = \sum (1/2) m \omega^2_i r_i^2$, where $\omega_i$ (with $i=1,2,3$) are the trapping frequencies along three spatial dimensions.

The local density is then
\begin{equation}\label{eq:nr}
    n(\mathbf{r})=\frac{g_{3/2}(\mathrm{e}^{(\mu-V(\mathbf{r}))/\kB T})}{\lambda^3} \; ,
\end{equation}
and the local phase space density $D(\mathbf{r})=n(\mathbf{r})\lambda^3$. The total number of particles in the excited states can be found by integrating over all space:
\begin{equation}\label{eq:Ntrap}
    N' =\int \frac{g_{3/2}(\mathrm{e}^{(\mu-V(\mathbf{r})/\kB T})}{\lambda^3} \, \mathrm{d}\mathbf{r} \; .
\end{equation}

The critical point for a trapped gas is the point at which the maximal local $D$ reaches the critical value of $\zeta(3/2)$. For a fixed $T$ it makes sense to define the critical point in terms of the critical total particle number $N_c$. For a harmonic trap, with the geometric mean of the three trapping frequencies $\bar{\omega}$, the integral in Eq.~(\ref{eq:Ntrap}) for $\mu=0$ gives
\begin{equation}\label{eq:Ncid}
\Ncid=\zeta(3)\left(\frac{\kB T}{\hbar \bar{\omega}}\right)^3 \; ,
\end{equation}
where $\zeta(3)\approx1.202$.
The equivalent expression for the transition temperature at a fixed particle number is given
by\footnote{Finite-size corrections slightly reduce the ideal-gas critical temperature, by $\kB \Delta T_c^0 =-\zeta(2)/(2 \zeta(3)) \, \hbar \omega_m\approx - 0.684 \, \hbar \omega_m$, where $\omega_m$ is the algebraic mean of the trapping frequencies \cite{Dalfovo:1999}.}
\begin{equation}\label{eq:Tcid}
\kB \To=\hbar \bar{\omega} \left(\frac{N}{\zeta(3)}\right)^{1/3}.
\end{equation}

\begin{figure}[t]
\centering
\includegraphics[width=0.65\columnwidth]{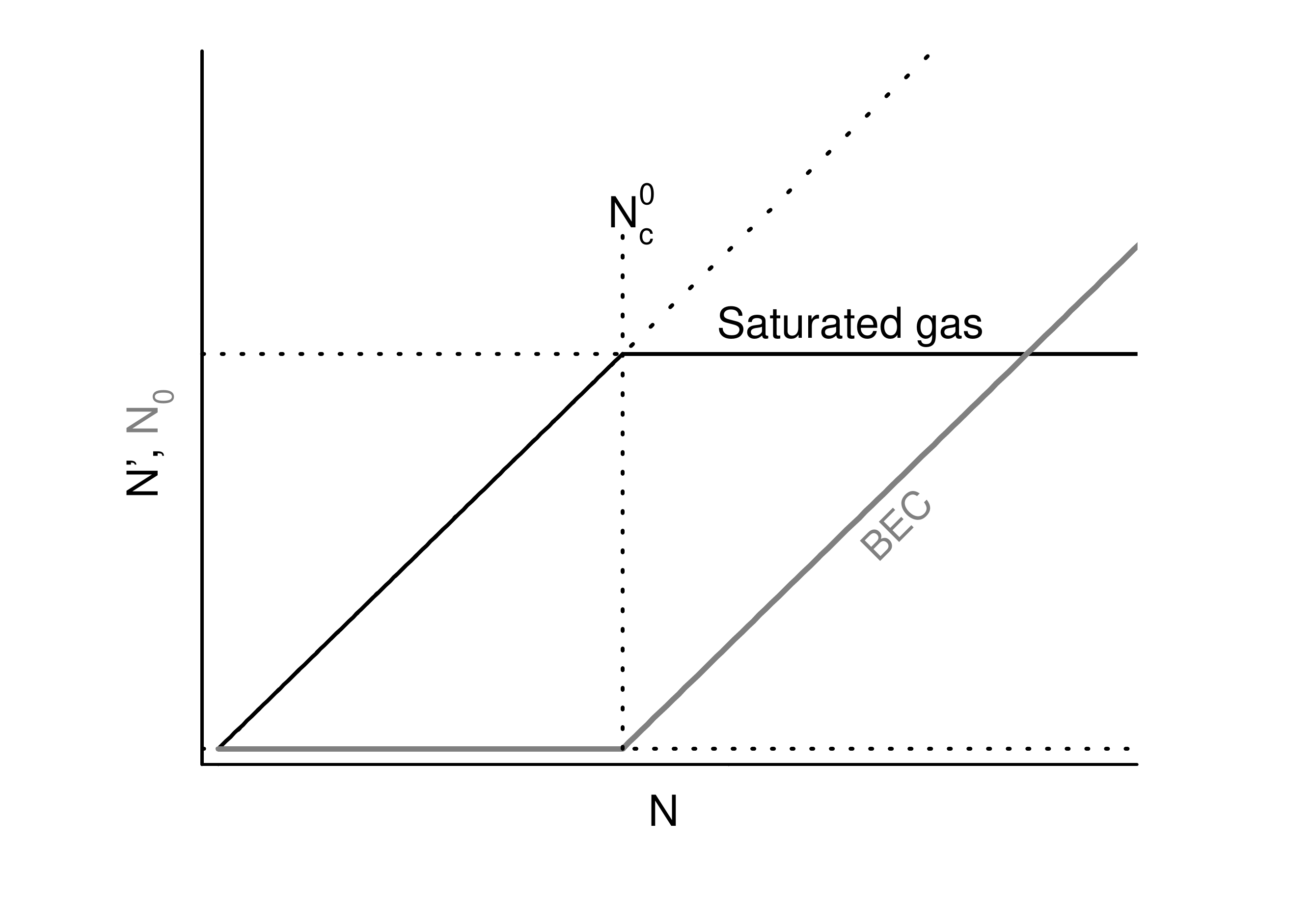}
\caption{Ideal Bose gas condensation. Number of thermal atoms $N'$ (black line) and number of condensed atoms $N_0$ (grey line) are plotted versus the total atom number $N$, at a fixed temperature. As atoms are added to trap  $N'=N$ and $N_0=0$ until the critical atom number $\Ncid$ is reached. At this point the excited states of the system saturate, and for $N> \Ncid$ we have $N'=\Ncid$ and $N_0=N-\Ncid$. }
\label{fig:intro_fig1}
\end{figure}

The ideal-gas picture of Bose-Einstein condensation driven by the purely statistical saturation of the excited states is simply summarised graphically in Fig. \ref{fig:intro_fig1}.
Here we plot the number of atoms  in the excited states, $N'$, and in the condensate, $N_0$, as the total atom number $N$ is increased at constant temperature. For $N < \Ncid$ no condensate is present and  $N' = N$. However for $N > \Ncid$ the thermal component is saturated at $N' = \Ncid$ and the number of condensed atoms is simply given by $N_0 = N - \Ncid$.
In sections \ref{sec:saturation} and \ref{sec:Tcshift} we will examine the effects of interactions both on the saturation of the excited states and on the value of the critical atom number.

\subsection{Interacting Bosons}
\label{sec:Interactions}

The dominant effects of interactions on Bose-Einstein condensation are quite different in a uniform system and in the experimentally pertinent case of a harmonically trapped atomic gas. This complex problem has a long history and for reviews we refer the reader to, for example, \cite{Dalfovo:1999,Pethick:2002,Pitaevskii:2003,Andersen:2004RMP}. Here we just briefly introduce some key points.

The simplest theoretical framework to address the effects of interactions in a Bose gas is the Hartree-Fock approximation \cite{Dalfovo:1999}. In this mean-field (MF) model one treats the thermal atoms as a ``noninteracting" gas of density $n'(\bf r)$ that experiences a self-consistent MF interaction potential  $g[2n_0({\bf r})+2n'({\bf r})]$, where $g=4 \pi \hbar^2 a /m$, $a$ is the s-wave scattering length, and $n_0(\bf r)$ the condensate density. We can then define an effective total potential
\begin{equation}
V_{\rm eff}({\bf r})=V({\bf r})+2g[n_0({\bf r})+n'({\bf r})] \; ,
\label{eq:Veff}
\end{equation}
and apply the LDA by replacing  $V({\mathbf{r}})$ with $V_{\mathrm{eff}}(\mathbf{r})$ in Eq.~(\ref{eq:nr}).
Meanwhile the condensed atoms feel an interaction potential $g[n_0({\bf r})+2n'({\bf r})]$, where the factor of two difference in the condensate self-interaction comes about due to the lack of the exchange interaction term for particles in the same state\footnote{This approach does not take into account the modification of the excitation spectrum  due to the presence of the condensate, which is included in more elaborate MF theories such as those of Bogoliubov \cite{Bogoliubov:1947} and Popov \cite{Popov:1987} (see also \cite{Dalfovo:1999}). However, it is often sufficient to give the correct leading order MF results.}.

In a uniform system, the MF potential gives just a spatially uniform energy offset and the most interesting effects arise due to beyond-MF quantum correlations.

On the other hand, in a harmonically trapped gas (with repulsive interactions) the inhomogeneous density results in a mean-field repulsion of atoms from the central high-density region. This geometrical effect often dominates and makes it harder to experimentally observe the more interesting beyond-MF physics.

\subsection{Chapter outline}

In section \ref{sec:experiment} we briefly outline our experimental procedure for performing precision measurements of the effects of interactions on Bose-Einstein condensation of an atomic gas.

In sections \ref{sec:saturation} and \ref{sec:Tcshift} we discuss the effects of interactions on the thermodynamics of a Bose gas with tuneable interactions. Here the range of interaction strengths we explore experimentally is such that the gas can always be assumed to be in thermal equilibrium. In section \ref{sec:saturation} we scrutinise the concept of saturation as the driving mechanism for Bose-Einstein condensation, and in section \ref{sec:Tcshift} we focus on the interaction shift of the critical point for condensation.  We compare the experimental results to both MF and beyond-MF theories and discuss how they relate to the case of a uniform Bose gas.

In section \ref{sec:nonequilibrium} we discuss the conditions required for equilibrium measurements, and non-equilibrium effects that are observed when they are violated.

\section{Precision measurements on a Bose gas with tuneable interactions }
\label{sec:experiment}

All the results presented here were obtained by performing conceptually simple experiments which are close in spirit to the ideal-gas theoretical plots of Fig. \ref{fig:intro_fig1}, but are performed for various strengths of repulsive interatomic interactions, characterised by the positive s-wave scattering length $a$.

Our experiments start with a partially condensed gas of $^{39}$K atoms,
produced in an optical dipole trap \cite{Campbell:2010}
with $\bar \omega/2\pi$ varying between 60 and 80\,Hz for data taken at different temperatures.
The strength of interactions in the gas can be tuned by applying a uniform external magnetic field in the vicinity of a Feshbach scattering resonance centred at 402.5 G \cite{Zaccanti:2009}.

In each experimental series we fix $a$ by choosing the value of the Feshbach field, and keep the temperature constant by fixing the depth of the optical trap. The total number of trapped atoms $N$ is then varied by holding the gas in the trap for a variable time $t$, up to tens of seconds. During this time $N$, initially larger than the critical value $\Nc$, slowly decays and eventually drops below $\Nc$. Meanwhile elastic collisions between the atoms act to redistribute the particles between the condensed and thermal components of the gas.

In each experimental run within a given series, corresponding to a particular hold-time $t$, the thermal atom number $N'$ and the condensate atom number $N_0$ are extracted from fits to the absorption images of the gas after $18-20\,$ms of free time-of-flight (TOF) expansion from the trap \cite{Ketterle:1999b,Gerbier:2004c}.
The interactions are rapidly turned off at the beginning of TOF, by tuning the Feshbach field to the $a=0$ point. This minimises the condensate expansion and allows us to home in on the critical point by reliably measuring condensed fractions as low as $0.1\,\%$.

For accurate measurements of the small interaction shift of the critical point it is particularly important to minimise various systematic errors, e.g. due to finite-size effects \cite{Dalfovo:1999}, uncertainties in the absolute calibration of $N$ and $\bar{\omega}$, and small anharmonic corrections to the trapping potential \cite{Campbell:2010}. We achieve this by performing ``differential measurements", always concurrently running two experimental series which are identical in every respect except for the choice of the scattering length \cite{Smith:2011}.

\section{Non-saturation of the excited states}
\label{sec:saturation}

\begin{figure}[t]
\centering
\includegraphics[width=0.65\columnwidth]{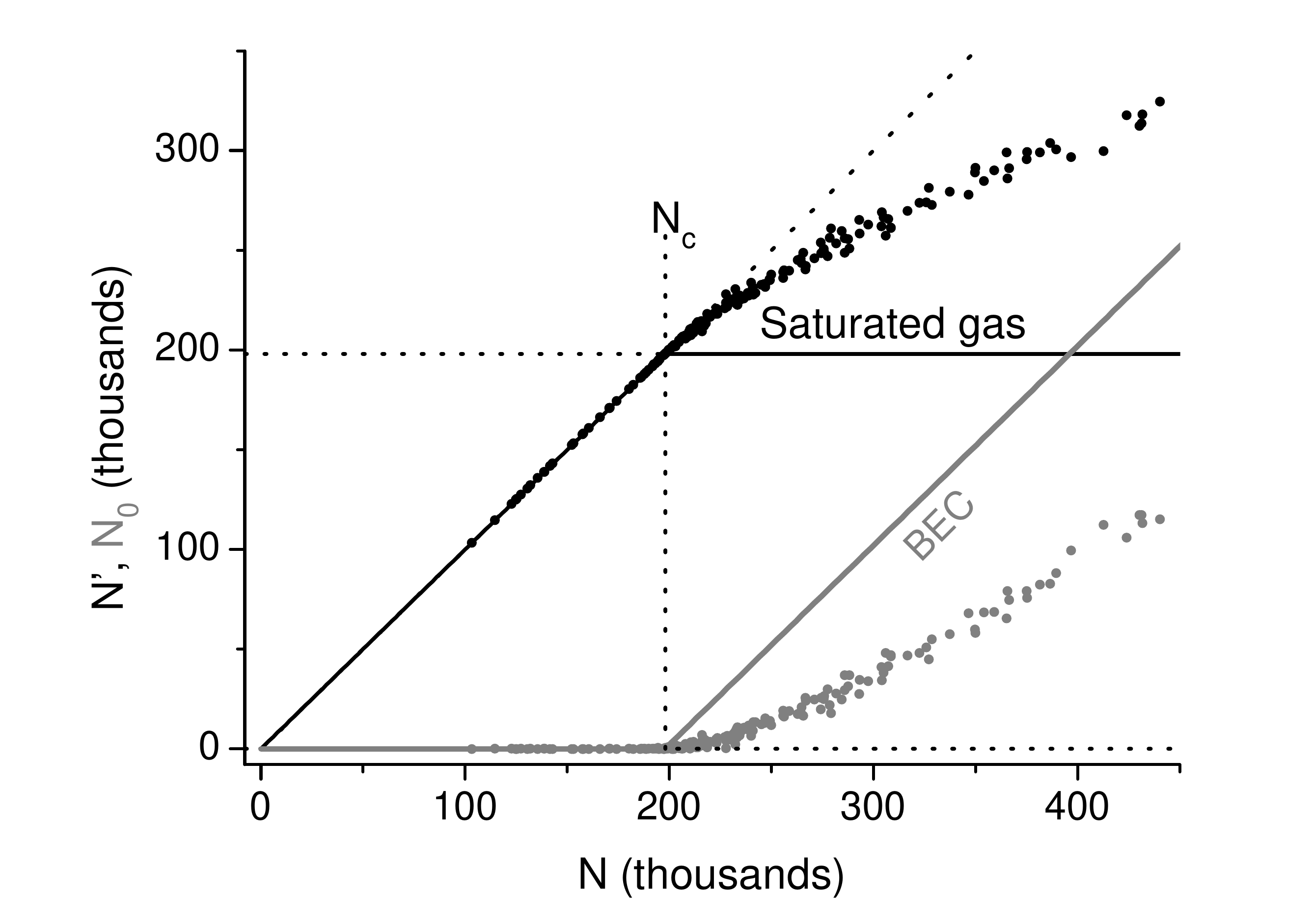}
\caption{Lack of saturation of the thermal component in a quantum degenerate atomic Bose gas.
$N'$ (black points) and $N_0$ (grey points) are plotted versus the total atom number $\Nt$
at $T=177$\,nK and $a=135\,a_0$. The corresponding predictions for a saturated gas are shown by black and grey solid lines. The critical point $\Nt=\Nc$ is marked by a vertical dashed line. (Figure adapted from \cite{Tammuz:2011}.)}
\label{fig:figure1}
\end{figure}

In this section, we focus on the concept of the saturation of the excited states as the underlying mechanism driving the BEC transition.
In superfluid $^4$He, which is conceptually associated with BEC, strong interactions preclude direct observation of purely statistical effects postulated by Einstein for an ideal gas. On the other hand it is generally accepted that a close-to-textbook BEC is observed in the weakly interacting atomic gases. Therefore, one might expect that the saturation inequality $N' \leq \Ncid$ is essentially satisfied in these systems, with just the value of the bound on the right-hand side slightly modified by interactions. However, as shown in Fig. \ref{fig:figure1}, this is far from being the case under typical conditions of an ultracold gas experiment.
Here, in an experimental series taken with $a=135\,a_0$ (where $a_0$ is the Bohr radius) and $T=177$\,nK, the measured critical atom number is
$\Nc \approx 200\,000$; if the total number of atoms is increased to $450\,000$, only half of the additional atoms accumulate in the condensate.

In order to explore the relationship between the experimentally observed non-saturation of the thermal component and the interatomic interactions, we first identify the relevant interaction energy. As a BEC is formed and then grows, the change in the average density of the condensed atoms is much larger than the change of the thermal density. Therefore one expects the non-saturation of the thermal component to result primarily from its interaction with the condensate (the $2gn_0(\mathbf{r})$ term in Eq.~(\ref{eq:Veff})). The relevant energy scale is then provided by \cite{Dalfovo:1999}
\begin{equation}
\mu_0=gn_0 (\mathbf{r}=0)=\frac{\hbar \bar \omega}{2}
\left(
15 N_0 \frac{a}{\aho}
\right)^{2/5}\ ,
\label{eq:mu}
\end{equation}
where $\aho=(\hbar/m\bar \omega)^{1/2}$ is the spatial extension of the ground state of the harmonic oscillator. The energy $\mu_0$ is the MF result for the chemical potential of a gas with $N_0$ atoms at zero temperature in the Thomas-Fermi limit  \cite{Dalfovo:1999}.

\begin{figure}[t]
\centering
\includegraphics[width=0.65\columnwidth]{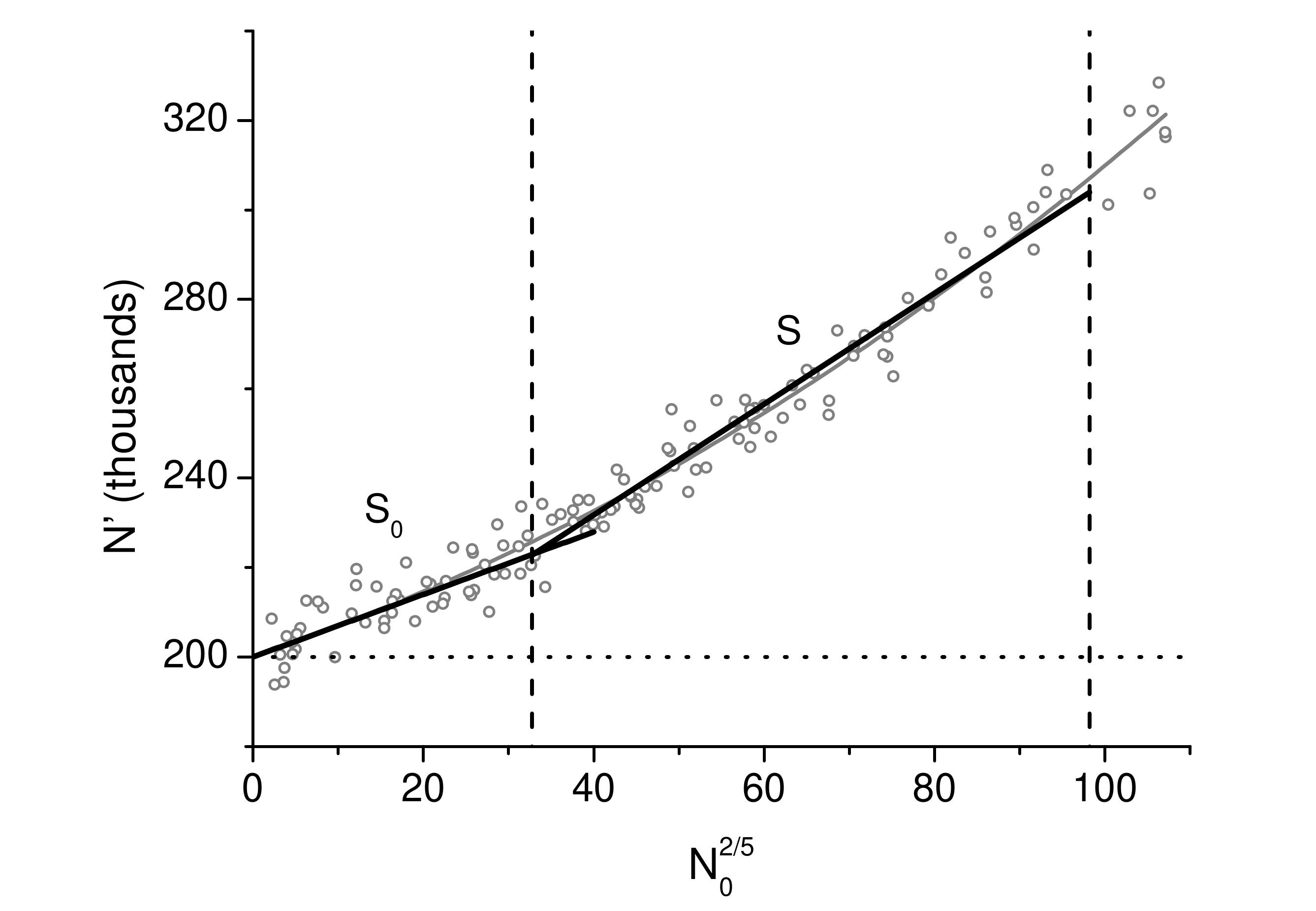}
\caption{Quantifying the lack of saturation. Here $N'$ is plotted as a function of $N_0^{2/5}$ for the same series as in Fig.~\ref{fig:figure1}. The horizontal dotted line is the saturation prediction $N'=\Nc$. The two black lines show the initial slope $S_0$ and the slope $S$ for $0.1<\mu_0\, /\,\kB T<0.3$.   The solid grey line is a guide to the eye based on a second-order polynomial fit. (Figure adapted from \cite{Tammuz:2011}.)}
\label{fig:TwoFifths}
\end{figure}

Guided by this scaling, for the data shown in Fig.~\ref{fig:figure1} we plot $N'$ as a function of $N_0^{2/5}$ in Fig.~\ref{fig:TwoFifths}. The growth of $N'$ with $N_0^{2/5}$ is not perfectly linear, so we quantify the non-saturation effect with two linear slopes: (1) the initial slope $S_0$ for $N_0 \rightarrow 0$, and (2) the course grained slope $S= \Delta[N']/\Delta [N_0^{2/5}]$ for $0.1<\mu_0\, /\,\kB T<0.3$ \cite{Tammuz:2011}.
The data shown in Fig.~\ref{fig:TwoFifths} can also be described by a second-order polynomial fit,
as described later.

The initial slope $S_0$ may be compared with the HF model.  In order to obtain the non-saturation effect to first order within the HF approach, we only consider the repulsive interaction of the thermal atoms with the condensate and not with other thermal atoms. From Eq.~(\ref{eq:Veff}) this leads to an effective
potential\footnote{Note that $gn_0(\mathbf{r})=\max\{\mu_0-V(r),0\}$.}
\begin{equation}
V_{\rm eff}({\bf r})=V({\bf r})+2gn_0({\bf r})=|V({\bf r})-\mu_0|+\mu_0\ .
\label{eq:Veffsat}
\end{equation}
Note that within this theory $N_c = \Ncid$ since $V_{\rm eff}({\bf r})=V({\bf r})$ when $N_0=0$. By integrating Eq.~(\ref{eq:Ntrap}) with the effective potential of Eq.~(\ref{eq:Veffsat}) one can predict a linear variation of $N'/N_c^0$ with the small parameter $\mu_0\, /\,\kB T$:
\begin{equation}
\frac{N'}{N_c^0}=1 + \alpha  \, \frac{\mu_0}{\kB T} \; ,
\label{eq:HF}
\end{equation}
with $\alpha = \zeta(2)/\zeta(3)\approx 1.37$. This first order non-saturation result is identical to that obtained in more elaborate MF approximations, which only modify higher order terms.

The origin of the non-saturation effect can be qualitatively understood by noting that interactions with the condensate modify the effective potential seen by the thermal atoms from a parabola into the ``Mexican hat" shape of Eq.~(\ref{eq:Veffsat}); this effectively allows the thermal component to occupy a larger volume, which grows with increasing $N_0$.

From Eqs.~(\ref{eq:mu}) and (\ref{eq:HF}) we define the HF non-saturation slope
\begin{equation} \label{eq:SHF}
S_{\rm HF} = \frac{dN'}{d (N_0^{2/5})} = \frac{\zeta(2)}{\zeta(3)} X
\end{equation}
where $X$ is the dimensionless parameter
\begin{equation}
X=\frac{\zeta(3)}{2}\left(\frac{\kB T}{\hbar \bar{\omega}}\right)^2 \left(\frac{15 \, a}{\aho}\right)^{2/5}.
\end{equation}
The measured $S_0$ is found to agree with $S_{\mathrm{HF}}$ for a range of $a$ and $T$ values \cite{Tammuz:2011}.

\begin{figure}[t]
\centering
\includegraphics[width=0.65\columnwidth]{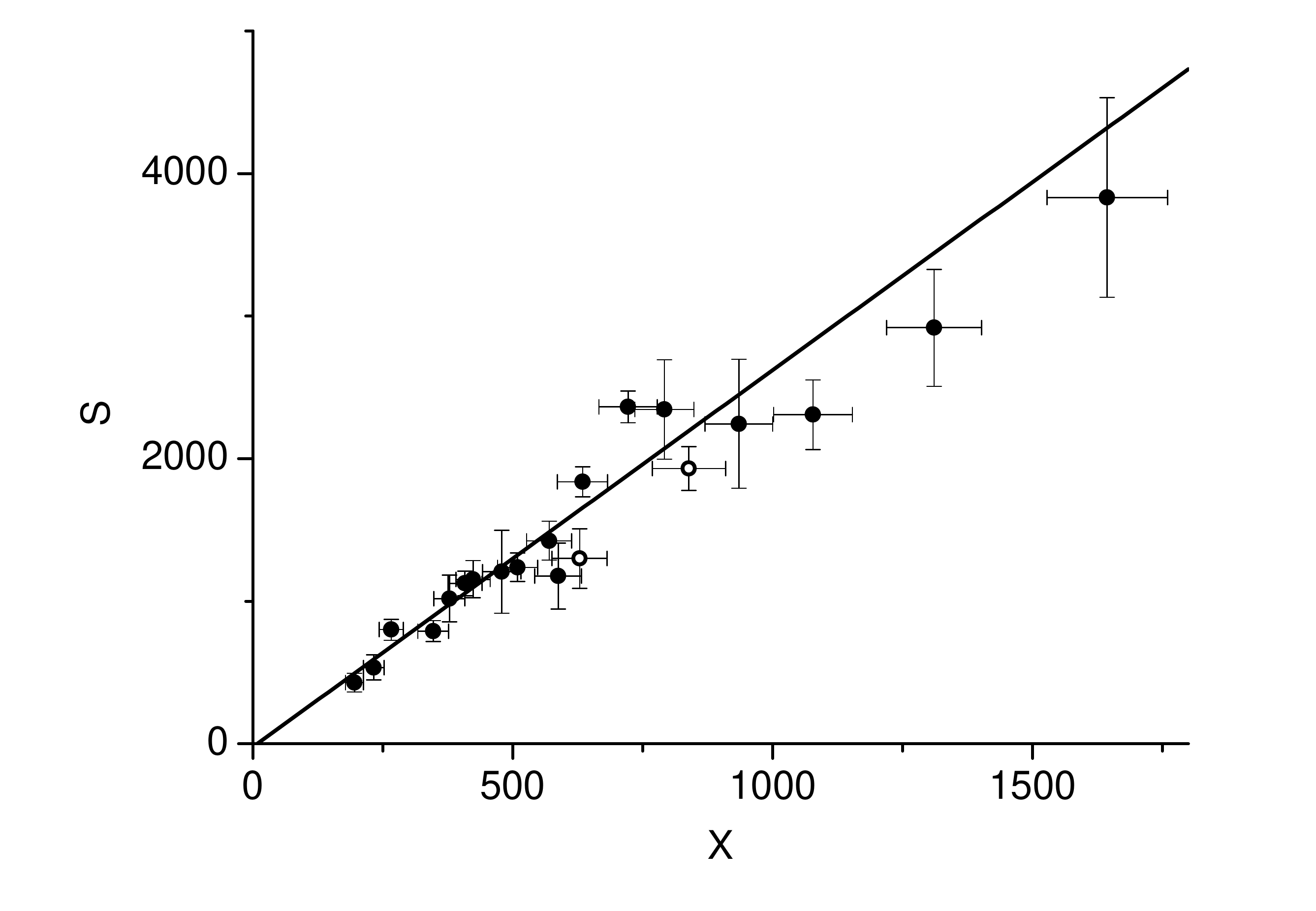}
\caption{Deviation from the saturation picture at a range of interaction strengths and temperatures.
The non-saturation slope $S$ is plotted versus the dimensionless interaction parameter $X \propto T^2 a^{2/5}$ (see text).  A linear fit (black line) gives $dS/dX = 2.6 \pm 0.3$\ and an intercept $S(0) = -20 \pm 100$, consistent with complete saturation in the ideal-gas limit. The data points are based on measurements with the $^{39}$K gas (closed circles) at a range of scattering lengths ($a = 40 - 356\,a_0$) and temperatures ($T = 115 - 284\,$nK), and two additional experimental series taken with a $^{87}$Rb gas (open circles). (Figure adapted from \cite{Tammuz:2011}.)
}
\label{fig:extrapolation}
\end{figure}

We now consider the non-saturation at higher $N_0$ values, where the data is not well described by Eq.~(\ref{eq:SHF}). Fig.~\ref{fig:extrapolation} summarises the non-saturation slopes $S(a,T)$ for a wide range of interaction strengths and temperatures. Within experimental error all data points fall onto a straight line with gradient $2.6\pm0.3$ and intercept $S(0) = -20 \pm 100$ when plotted against the dimensionless interaction parameter $X$.

The first and most important thing to notice is that both non-saturation slopes, $S_0$ and $S$, tend to zero for $X \rightarrow 0$.
These experiments thus confirm the concept of a saturated Bose gas, and Bose-Einstein condensation as a purely statistical phase transition in the non-interacting limit.

A question that is still open is the deviation of $S$ from the first order HF result $S_{\mathrm{HF}}$, i.e. $dS/dX \approx 2.6$ versus $\zeta(2)/\zeta(3) \approx 1.37$ predicted from HF theory.  This discrepancy can partially be explained by higher order terms in the mean-field theory, either directly from using Eq.~(\ref{eq:Veffsat}) in Eq.~(\ref{eq:Ntrap}), or using more elaborate MF theories such as the Popov approximation \cite{Pethick:2002}. However the effect is far stronger in the experimental results than any of these MF theories predict. To see this we consider the next order term in Eq.~(\ref{eq:HF}), writing $N'/\Ncid=1+\alpha (\mu_0/\kB T)+\alpha_2 (\mu_0/\kB T)^2$ with $\alpha=1.37$.
Experimentally, by identifying $S$ with the gradient of this quadratic function evaluated at $\mu_0/\kB T= 0.2$,  we get $\alpha_2=3\pm0.7$. For comparison the Popov approximation gives $\alpha_2 \approx 0.6$. At present the reason for this discrepancy and the possible role of beyond-MF effects are unclear, and require further investigation.

In summary, we can quantify the non-saturation of the thermal component in a harmonically trapped gas by writing the number of thermal atoms in a partially condensed cloud as
\begin{equation}
N'=N_c+S_0 N_0^{2/5} + S_2 N_0^{4/5} \; ,
\end{equation}
where
\begin{equation}\label{eq:S0}
S_0=\frac{\zeta(2)}{2}\left(\frac{\kB T}{\hbar \bar{\omega}}\right)^2 \left(\frac{15a}{\aho}\right)^{2/5}
\end{equation}
and
\begin{equation}\label{eq:S2}
S_2=(3 \pm0.7) \, \frac{\zeta(3)}{4} \, \frac{\kB T}{\hbar \bar{\omega}} \, \left(\frac{15a}{\aho}\right)^{4/5}.
\end{equation}

We have seen that in a harmonic trap the dominant non-saturation effect is ``geometric", arising from an interplay of the mean-field repulsion and the inhomogeneity of the condensate density. It is then interesting to consider the case of a uniform system, where this geometric effect is absent. Within MF theory, as the total density of a uniform Bose gas is increased past the critical value the thermal density $n'$ actually decreases. This is due to the fact that the atoms in the condensate have less interaction energy, as discussed in section \ref{sec:Interactions}. In addition, close to the transition, beyond-MF effects are expected to play an important role. This would make measurements in a uniform system (or of the local density in a trapped system) particularly interesting.

\section{Interaction shift of the transition temperature}
\label{sec:Tcshift}

Having considered the effect of interactions on the saturation of the thermal component we now consider the location of the critical point itself.

It is generally accepted that in a uniform system there is no interaction shift of the critical temperature $T_c$ at the level of mean-field theory. However, the beyond-MF correlations between particles which develop near the critical point are expected to shift $T_c$
\cite{Lee:1957TcShift, Lee:1958TcShift2,Bijlsma:1996, Baym:1999, Holzmann:1999, Reppy:2000, Arnold:2001, Kashurnikov:2001, Holzmann:2001, Baym:2001, Kleinert:2003, Andersen:2004RMP, Holzmann:2004}.
For several decades there was no consensus on the functional form, or even on the sign of this $T_c$ shift (for an overview see  e.g. \cite{Arnold:2001, Baym:2001, Andersen:2004RMP, Holzmann:2004}). It is now generally believed that the shift is positive and to leading order
given by \cite{Arnold:2001, Kashurnikov:2001}:
\begin{equation}
\frac{\dT}{\To}  \approx 1.3 \, a n^{1/3} \approx 1.8  \, \frac{a}{\lo} \, ,
\label{eq:uniformTcShift}
\end{equation}
where $\dT = T_c - \To$  and $\lo$ is the thermal wavelength at temperature $\To$. Equivalently, the $n_c$ shift at constant $T$ is $\Delta n_c/n_c^0 \approx - (3/2) \dT /\To$. The positive $\dT$ implies that condensation occurs at a phase space density below the ideal-gas critical value of 2.612.

\begin{figure}[t]
\centering
\includegraphics[width=0.65\columnwidth]{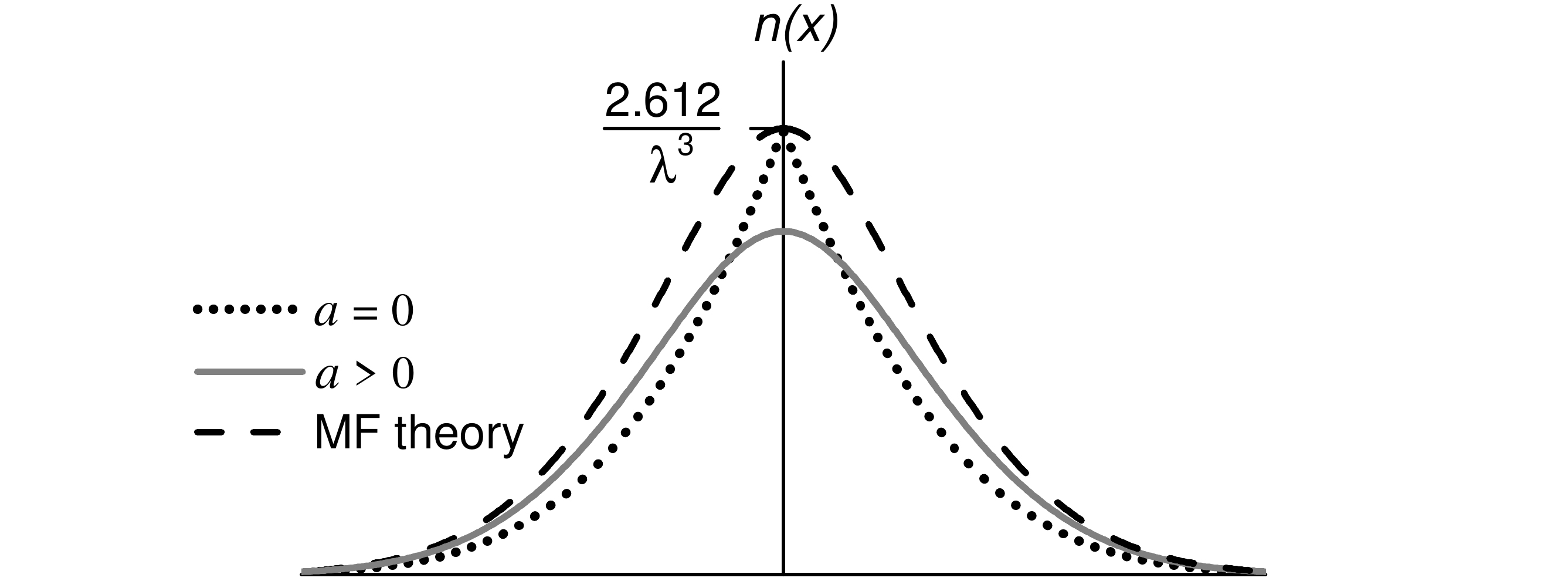}
\caption{Opposing effects of interactions on the critical point of a trapped Bose gas. We sketch the density distribution in a harmonic potential $V(r)$ at the condensation point. Compared to an ideal gas (dotted line) at the same temperature, repulsive interactions reduce the critical density, but also broaden the density distribution (solid line). Mean-field theory (dashed line) captures only the latter effect, and predicts an increase of the critical atom number $N_c$ at fixed $T$, equivalent to a decrease of $T_c$ at fixed $N$. (Figure adapted from \cite{Smith:2011}.)}
\label{fig:TcCartoon}
\end{figure}

The problem of the $T_c$ shift in a harmonically trapped gas is even more complex. In this case, at least for weak interactions, the shift is dominated by an opposing effect that reduces the critical temperature \cite{Giorgini:1996}.
This negative $T_c$ shift is due to the broadening of the density distribution by repulsive interactions (see Fig.~\ref{fig:TcCartoon}). To leading order it can be calculated analytically using MF theory \cite{Giorgini:1996}, by self-consistently solving Eq.~(\ref{eq:nr}) with $V_{\mathrm{eff}}=V(\mathbf{r})+2gn'(\mathbf{r})$:
\begin{equation}
\frac{\dT}{\To} \approx - 3.426 \, \frac{a}{\lo}  \; .
\label{eq:StringariMF}
\end{equation}
For the experimentally relevant range of interaction strengths, $0<a/\lo<0.05$, we numerically obtain the second-order MF shift, $\approx 11.7 \, (a/\lo)^2$ \cite{Smith:2011b}.

The two opposing effects of repulsive interactions on the critical point of a trapped gas are visually summarised in Fig.~\ref{fig:TcCartoon}, where we sketch the density distribution at the condensation point for an ideal and an interacting gas at the same temperature\footnote{The shift of the critical point can be equivalently expressed as $\Delta T_c(N)$ or $\Delta N_c(T)$, with $\Delta N_c(T)/\Ncid \approx -3 \dT /\To$.}.
In the spirit of LDA, the critical density should be reduced by interactions. However, interactions also broaden the density distribution.   For weak interactions the latter effect is dominant, making the overall interaction shift $\Delta N_c(T)$ positive, or equivalently $\Delta T_c(N)$ negative.

The dominance of the negative MF shift of $T_c$ over the positive beyond-MF one goes beyond the difference in the numerical factors in Eqs.~(\ref{eq:uniformTcShift}) and (\ref{eq:StringariMF}). In a harmonic trap, at the condensation point only the central region of the cloud is close to criticality\footnote{The size of the central critical region is $r_c \sim (a/\lo) R_T$, where $R_T = \sqrt {\kB T / m\omega^2}$ is the thermal radius of the cloud \cite{Arnold:2001b}.}; this reduces the net effect of critical correlations so that they affect $T_c$ only at a higher order in $a/\lo$. The MF result of Eq.~(\ref{eq:StringariMF}) should therefore be exact at first order in $a/\lo$.
The higher-order beyond-MF shift is still expected to be positive, but the theoretical consensus on its value has not been reached \cite{Houbiers:1997, Holzmann:1999b, Arnold:2001b, Davis:2006, Zobay:2009}.

\subsection{Measurements on a harmonically trapped Bose gas}

Since the early days of atomic BECs there have been several measurements of the interaction $T_c$ shift in a harmonically trapped gas \cite{Ensher:1996, Gerbier:2004,Meppelink:2010}. These experiments,
performed at $a/\lo $ ranging from $0.007$ \cite{Meppelink:2010} to  0.024 \cite{Gerbier:2004},
nicely confirmed the theoretical prediction for the linear MF shift of Eq.~(\ref{eq:StringariMF}), but could not discern the beyond-MF effects of critical correlations.

\begin{figure} [t]
\centering
\includegraphics[width=1.0\columnwidth]{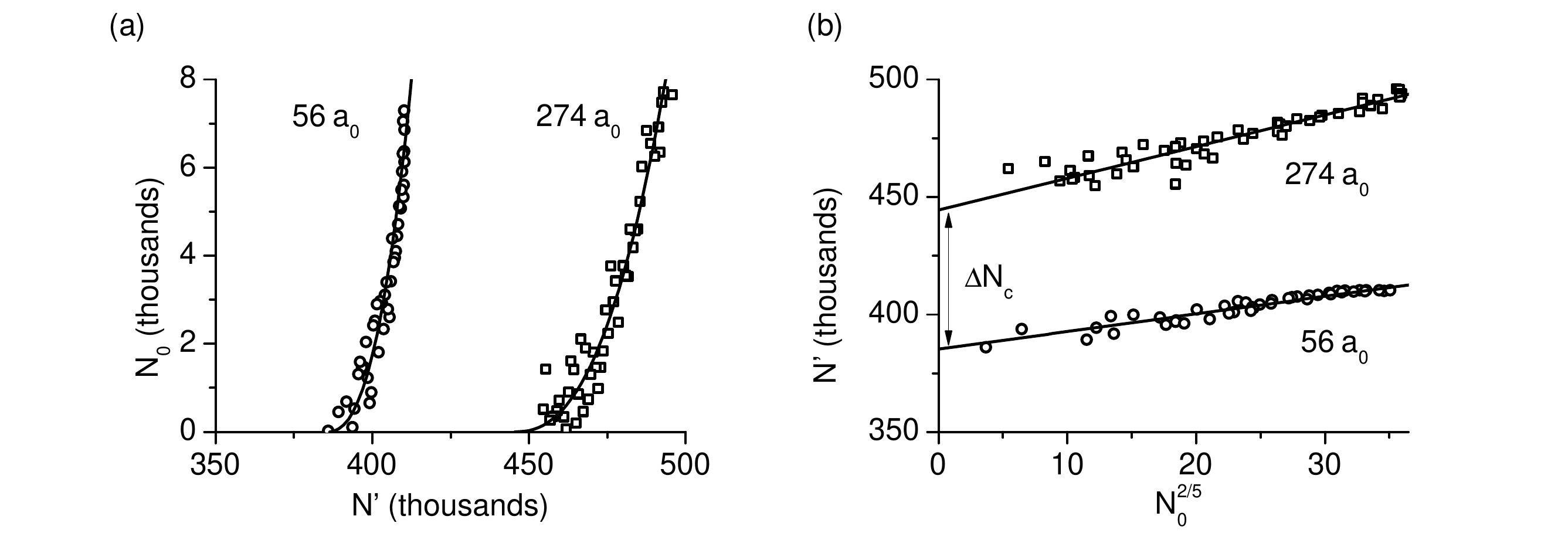}
\caption{Determination of the critical point and the differential interaction shift.  (a) Condensed ($N_0$) versus thermal ($N'$) atom number for two concurrently taken data series with $a = 56\,a_0$ (circles) and $a=274\,a_0$ (squares). Note that all points correspond to  condensed fractions below $2\,\%$. The data is scaled to the same temperature ($T=240\,$nK) and shows the shift of the critical point in the form $\Delta N_c(T)$. Solid lines show the extrapolation to $N_0=0$, necessary  to accurately determine $N_c$. (b) $N'$ is plotted versus $N_0^{2/5}$ for the same data as in (a), showing more clearly the extrapolation procedure. (Figure adapted from \cite{Smith:2011}.)}
\label{fig:Nc}
\end{figure}

The recent measurements \cite{Smith:2011} presented here provided the first clear observation of the beyond-MF $T_c$ shift in a trapped atomic gas. Several improvements contributed to making this possible. First, we explored slightly higher interaction strengths, up to $a/\lo \approx 0.04$. Second, by performing precision measurements outlined in section \ref{sec:experiment}, and directly accessing the small differential $T_c$ shift due to the variation in $a/\lo$, we significantly reduced the experimental error bars. Third, understanding the non-saturation effects discussed in section \ref{sec:saturation} was also essential for accurately determining the critical point from the measurements performed {\it close} to it (see Fig.~\ref{fig:Nc}).

Fig.~\ref{fig:Nc} illustrates a differential measurement with $a=56\,a_0$ and $a = 274\,a_0$.
The rise of $N_0$ in Fig.~\ref{fig:Nc}(a) is not simply vertical because the thermal component is not saturated at $N_c$ (see section \ref{sec:saturation}). It is therefore essential to carefully extrapolate $N'$ to the $N_0=0$ limit in order to accurately determine $N_c$. For small $N_0$ the extrapolation is done using $N'=N_c+S_0 N_0^{2/5}$, with the non-saturation slope $S_0(T,\bar{\omega},a)$ given by Eq. (\ref{eq:S0}).

\begin{figure} [t]
\centering
\includegraphics[width=0.65\columnwidth]{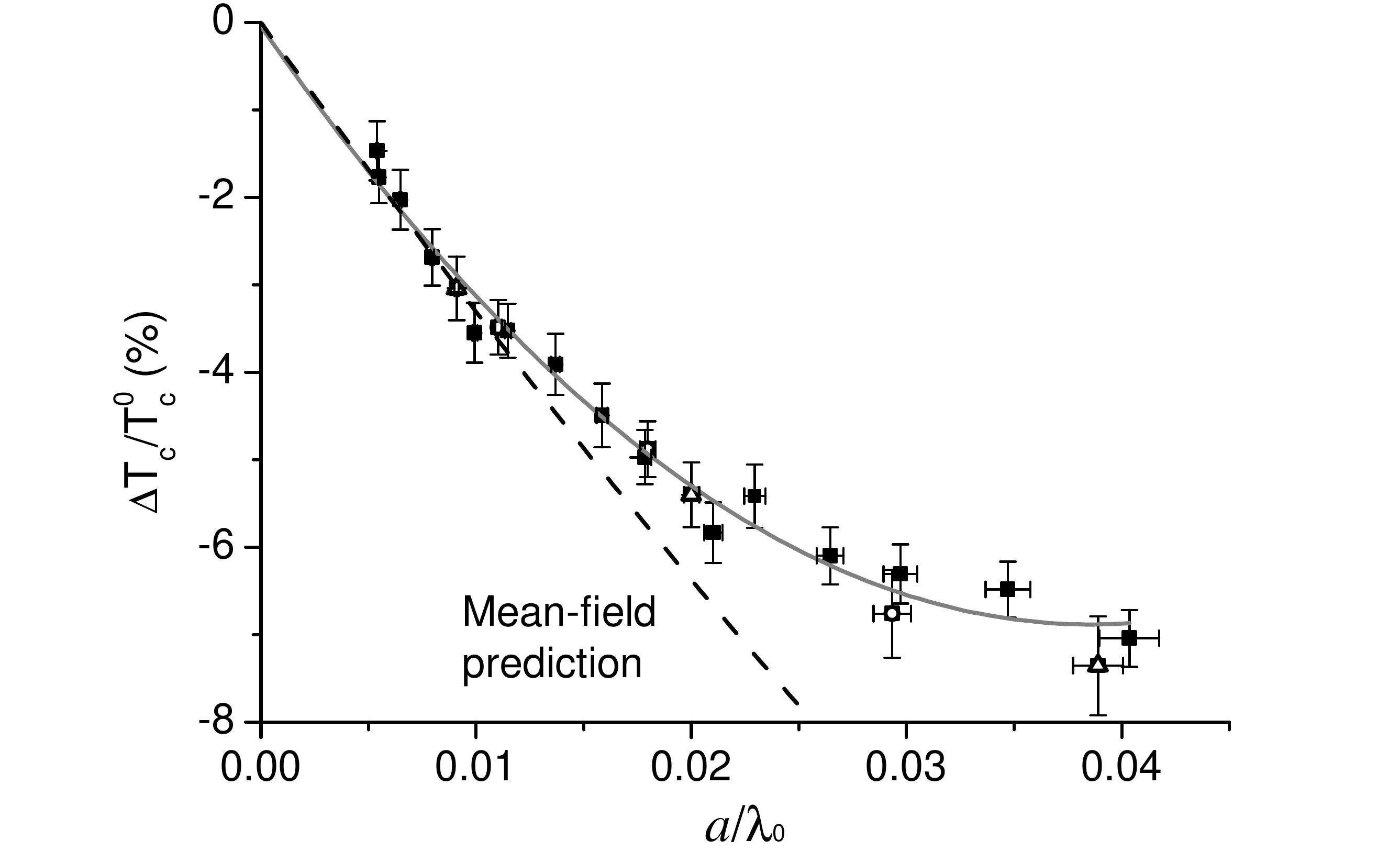}
\caption{Interaction shift of the critical temperature. Data points were taken with $N \approx 2 \times 10^5$ (open circles), $4 \times 10^5$ (black squares), and $8 \times 10^5$ (open triangles) atoms. The dashed line is the MF prediction. The solid line shows a second-order polynomial fit to the data (see text). Vertical error bars are statistical, while systematic errors in $\Delta T_c / \To$ are assessed to be $<1\,\%$ \cite{Smith:2011}.  (Figure adapted from \cite{Smith:2011}.)}
\label{fig:TcShift}
\end{figure}

In Fig.~\ref{fig:TcShift} we summarise our measurements of  $\Delta T_c / \To$ \cite{Smith:2011}. The data taken with different atom numbers, $N \approx (2-8) \times 10^5$, fall onto the same curve, confirming that the results depend only on the interaction parameter $a/\lo$.
The MF prediction agrees very well with the data for $a/\lo \lesssim 0.01$, but for larger $a/\lo$ there is a clear deviation from this prediction. All the data are fitted well by a second-order polynomial
\begin{equation}
\frac{\dT}{\To} \approx b_1 \, \frac{a}{\lo} + b_2 \left( \frac{a}{\lo} \right)^2\, ,
\label{eq:TrapTc}
\end{equation}
with $b_1= - 3.5 \pm 0.3$ and $b_2 = 46 \pm 5$. The value of $b_1$ is in agreement with the MF prediction of $-3.426$. The $b_2$ value strongly excludes the MF result of $b_2^{\rm MF} \approx 11.7$ and its sign is consistent with the expected effect of beyond-MF critical correlations.
Fixing $b_1=-3.426$ (which is expected to be exact even including beyond-MF effects) gives an improved estimate of $b_2=42\pm2$.

\subsection{Connection with a uniform Bose gas}

In order to make a connection between the experiments on trapped atomic clouds and the theory of a uniform Bose gas we also need to consider the effect of interactions on the critical chemical potential $\mu_c$.

In a uniform gas the interactions differently affect $T_c$ (or equivalently $n_c$) and $\mu_c$ at both MF and beyond-MF level. The simple MF shift $\beta \mumf = 4 \, \zeta(3/2) \, a/\lo$ (where $\beta = 1/\kB T$) has no effect on condensation. To lowest beyond-MF order we have\footnote{Note that $B_2$ is not just a constant but contains logarithmic corrections in $a/\lo$ \cite{Arnold:2001}. We neglect these in our discussion since they are not discernible at the current level of experimental precision.}:
\begin{equation}
\beta \mu_c \approx \beta \mumf + B_2 \left( \frac{a}{\lo} \right)^2 \; .
\label{eq:Muc}
\end{equation}
We see that there is a qualitative difference between Eqs. (\ref{eq:uniformTcShift}) and (\ref{eq:Muc}). Specifically, we have $\nmf - n_c \propto a/\lo$, but $\mumf - \mu_c \propto (a/\lo)^2$. This difference highlights the fact that the problem of the $T_c$ shift is non-perturbative; near criticality the equation of state does not have a regular expansion\footnote{The non-interacting equation of state (Eq.~(\ref{eq:psd})) cannot be expanded about $D_c$ in $\beta \mu$, but rather in $\sqrt{-\beta \mu}$; up to first order this expansion gives  $D=D_c-2\sqrt{\pi}\sqrt{-\beta \mu}$. This scaling goes some way in explaining the qualitative difference between Eqs. (\ref{eq:uniformTcShift}) and (\ref{eq:Muc}) although it cannot be used quantitatively.}
in $\mu$, otherwise one would get $\Delta n_c \propto \mu_c - \mumf$.

\begin{figure} [t]
\centering
\includegraphics[width=0.65\columnwidth]{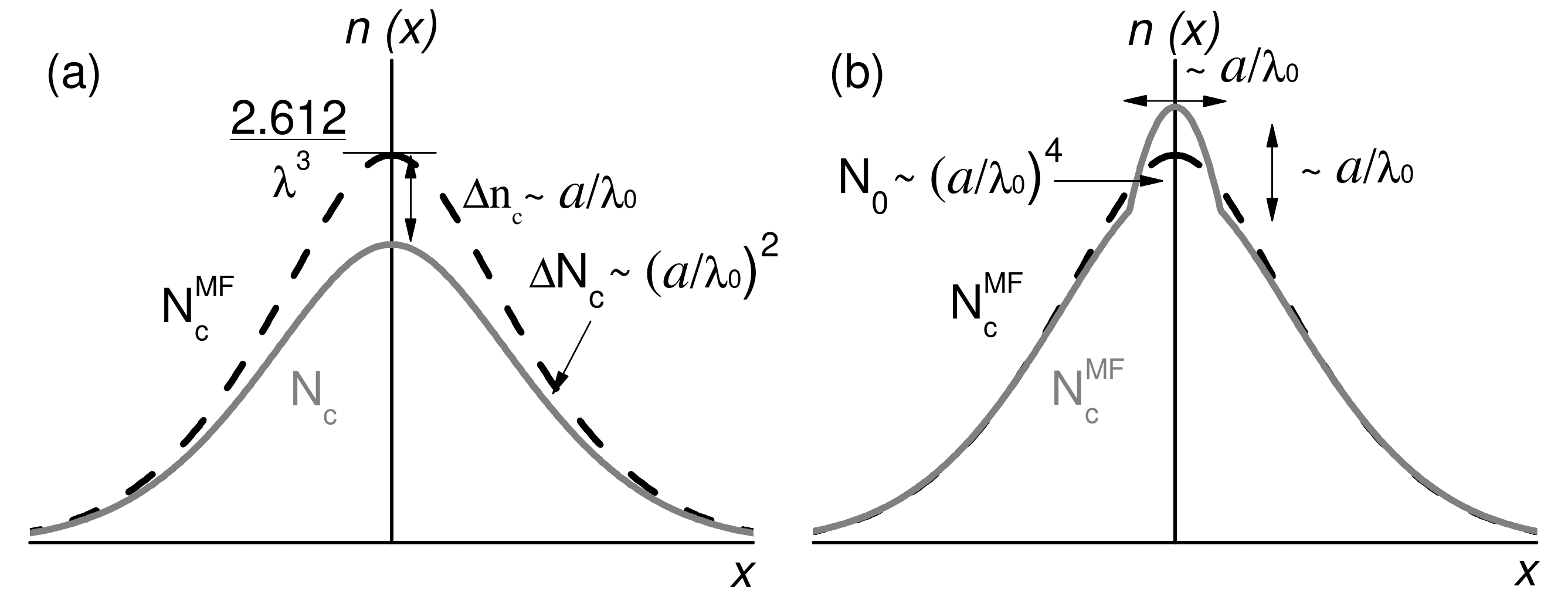}
\caption{ Beyond-MF effects near the critical point in a harmonically trapped Bose gas. (a) For a fixed $T$, the density distribution at the critical point $N=N_c < \Nmf$ (solid grey line) is compared with the MF prediction (dashed line). In the trap centre we expect $\nmf - n_c  \propto a/\lo$, characteristic of a uniform system. However the experimentally measured $\Nmf - N_c \propto (a/\lo)^2$ is dominated by the density shift {\it outside} the central critical region,
and is not directly related to the $n_c$ shift.
(b) If $N$ is increased to $\Nmf > N_c$, a condensate induced by critical correlations forms within the critical region of size $\propto a/\lo$. The condensed atom number  $N_0 \propto (a/\lo)^4$ directly relates to the critical density shift $\Delta n_c \propto a/\lo$. (Figure adapted from \cite{Smith:2011b}.)}
\label{fig:Cartoon}
\end{figure}

For a harmonic trap, within LDA the uniform-system results for $n_c$ and $\mu_c$ apply in the centre of the trap, and elsewhere local $\mu$ is given by Eq.~(\ref{eq:mulocal}). The result for the $T_c$ shift however does not carry over easily to the non-uniform case.
As illustrated in Fig.~\ref{fig:Cartoon}(a), in the centre of the trap we expect $\Delta n_c \propto a/\lo$, but the measured beyond-MF $\Delta N_c \propto (a/\lo)^2$ (see Eq.~(\ref{eq:TrapTc})).

In fact, the experimentally observed $T_c$ shift (Eq.~(\ref{eq:TrapTc})) qualitatively mirrors Eq.~(\ref{eq:Muc}), with the MF term linear in $a/\lo$ and the beyond-MF one quadratic in $a/\lo$ (to leading orders).
This similarity can be understood as follows: The interaction shift of $\mu_c$ affects the density everywhere in the trap; outside the small critical region the equation of state is regular in $\mu$ and the local density shift is simply proportional to the local $\mu$ shift; the contribution to the total $N_c$ shift from the non-critical region greatly outweighs the contribution from within the critical region and therefore one qualitatively expects $N_c - \Nmf \propto \mu_c - \mumf$. More quantitatively, this connection is given by\footnote{Note that this relationship is closely related to Eq.~(\ref{eq:HF}).} \cite{Arnold:2001}
\begin{equation}
\label{eq:B2}
-3(b_2-b_2^{MF})=\frac{\zeta(2)}{\zeta(3)}B_2 \; .
\end{equation}

The key conclusion of this discussion is therefore that the beyond-MF $T_c$ shift observed in a trapped gas is directly related to the beyond-MF $\mu_c$ shift (in either trapped or uniform system). It does not however directly reveal the expected linear $n_c$ shift and the theoretically most intriguing non-perturbative connection between $\mu_c$ and  $n_c$ shifts. In fact one could say that the historically puzzling (theoretical) connection between $n_c$ and $\mu_c$ is just replaced by the puzzling connection between (expected) $n_c$ and (measured) $N_c$.

This problem can be partly overcome by studying the condensed fraction $f_0$ in a trapped atomic cloud at the MF-predicted critical point \cite{Smith:2011b}. By definition $f_0$ vanishes within MF theory, and so directly measures the effect of critical correlations which shift $T_c$ above $\Tmf$.
At a fixed $T$ we consider the condensed fraction of a gas, $N_0/N$, at the point where $ N= \Nmf > N_c$, as illustrated in Fig.~\ref{fig:Cartoon}(b).
The analogous quantity for a uniform gas was first theoretically studied by Holzmann and Baym \cite{Holzmann:2003}, who showed
that\footnote{This scaling holds for any distance from the critical point given by $(\mu - \mu_c)(\lo/a)^2 = {\rm const}$. By applying it to the MF critical point we neglect the logarithmic corrections to $\mumf - \mu_c$, which are so far not experimentally observable.}
$n_0/n \propto \Delta n_c \propto a/\lo$.
The condensate density $n_0$ vanishes at the point where the local $\mu = \mu_c$, so from Eqs.~(\ref{eq:mulocal}) and (\ref{eq:Muc}) we get that the spatial extension of the condensate is also $\propto a/\lo$, and hence $f_0 \propto (a/\lo)^4$.

\begin{figure} [t]
\centering
\includegraphics[width=0.65\columnwidth]{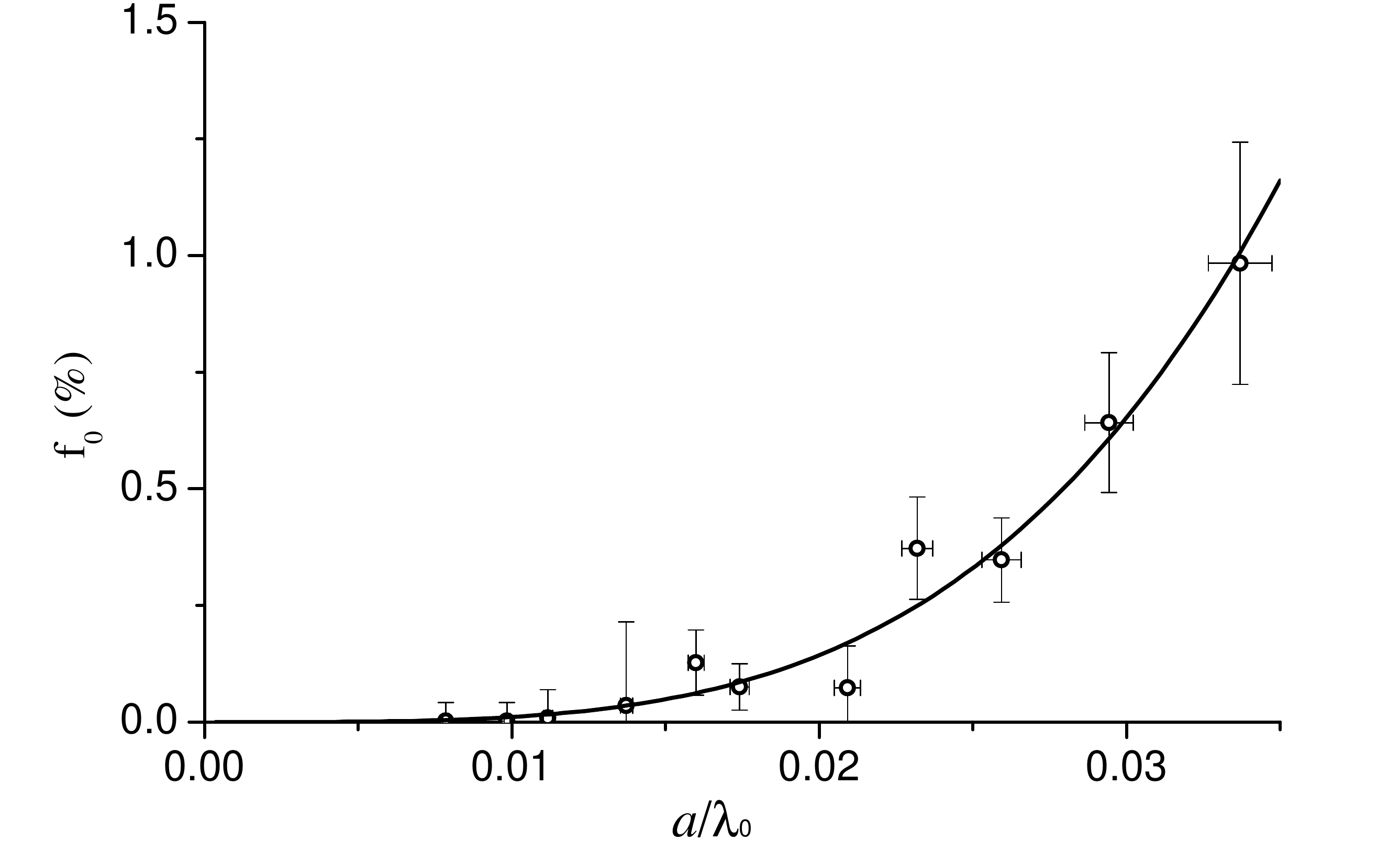}
\caption{ Condensed fraction of an atomic gas induced by critical correlations. The condensed fraction $f_0 = N_0/N$ is measured for $N= \Nmf > N_c$. A fit to the data (solid line) with the function $f_0 \propto (a/\lo)^x$ gives an exponent $x=3.9 \pm 0.4$, in agreement with the predicted $x = 4$. (Figure adapted from \cite{Smith:2011b}.)}
\label{fig:quartic}
\end{figure}

The quartic scaling of $f_0$ with $a/\lo$ is thus directly related to the expected linear scaling of $\Delta n_c$ with $a/\lo$; the two scalings are simply connected by the volume of the critical region, $\propto (a/\lo)^3$. This scaling is indeed confirmed experimentally \cite{Smith:2011b}, as shown in Fig.~\ref{fig:quartic}.

One can take the comparison with theory beyond just the scaling of $f_0$ with $a/\lo$, and quantitatively compare the measured $N_0$ at the MF critical point with the Monte-Carlo (MC) calculations \cite{Prokofev:2004} for a uniform gas.
This also works out very well, with the measured and predicted $N_0$ agreeing within a few percent \cite{Smith:2011b}. It is however important to carefully summarise the conceptual steps involved in this comparison:

(i) On the one hand, the measurements of the $T_c$ shift \cite{Smith:2011} experimentally provide (up to logarithmic corrections) the value of $\mumf - \mu_c \propto (a/\lo)^2$, via Eqs.~(\ref{eq:TrapTc}), (\ref{eq:Muc}), and (\ref{eq:B2}) \cite{Smith:2011b}.

(ii) On the other hand, the MC calculations \cite{Kashurnikov:2001, Prokofev:2004} that predict the $n_c$ shift of Eq.~(\ref{eq:uniformTcShift}) also provide tabulated values of the uniform-system condensate density $n_0$ for any $\mu-\mu_c \propto (a/\lo)^2$.

(iii) Combining these two results and the LDA (Eq.~(\ref{eq:mulocal})) we calculate the expected $N_0$ in a trapped gas with $N = \Nmf$ and find excellent agreement with the measurements shown in Fig.~\ref{fig:quartic} \cite{Smith:2011b}.

Overall this provides strong cross-validation of theory and experiment. However it is important to note that the two different measurements of beyond-MF effects (i.e. of the $T_c$ shift and $f_0$) do not provide two \textit{independent} quantitative tests of the uniform-system theory. Instead, what we have shown is that they are consistently connected via the MC calculations for a uniform system. Finally it is also important to stress that we are still lacking a direct measurement of the $n_c$ shift, which would explicitly test the historically most debated theoretical result of Eq.~(\ref{eq:uniformTcShift}).
This goal remains open for future measurements, either of the local density in a harmonically trapped gas or on a uniformly trapped atomic gas.

\section{Equilibrium criteria and non-equilibrium effects}
\label{sec:nonequilibrium}

Finally, we discuss the criteria for the measurements on a trapped Bose gas to faithfully represent its equilibrium properties, and the non-equilibrium effects revealed when they are violated. It is helpful to distinguish two types of non-equilibrium behaviour, transient and intrinsic.

Transient non-equilibrium effects are more familiar, and occur whenever some system parameter, such as the interaction strength, is rapidly changed (quenched). After such a quench the system relaxes towards its new equilibrium.
Classically, one can estimate the relaxation time to be several elastic scattering times $1/\gamma_{\rm el}$, where $\gammael$ is the elastic scattering rate \cite{Monroe:1993, Arndt:1997, Newbury:1995, Kavoulakis:2000c}.

However a system with continuous dissipation can only be ``close to" thermodynamic equilibrium and is to some extent always intrinsically out of equilibrium. The proximity to equilibrium broadly depends on the competition between relaxation and dissipation. For an atomic gas, this leads to a criterion based on the dimensionless parameter $\gamma_{\rm el} \tau$, where $\tau$ is some characteristic dissipation time, e.g.\ for atom loss. In practice, the relevant $\tau$ and the criteria for equilibrium measurements depend on the required measurement precision.

In the case of the $T_c$ measurements presented in section \ref{sec:Tcshift}, $N_c$ is determined to about $1\,\%$, so we require that the gas continuously (re-)equilibrates on a timescale $\tau$ corresponding to only $1\,\%$ atom-loss. This requires about 100 times higher $\gamma_{\rm el}$ than one would naively conclude by taking the $1/e$ lifetime of the cloud as the relevant timescale.

\begin{figure}[t]
\centering
\includegraphics[width=0.65\columnwidth]{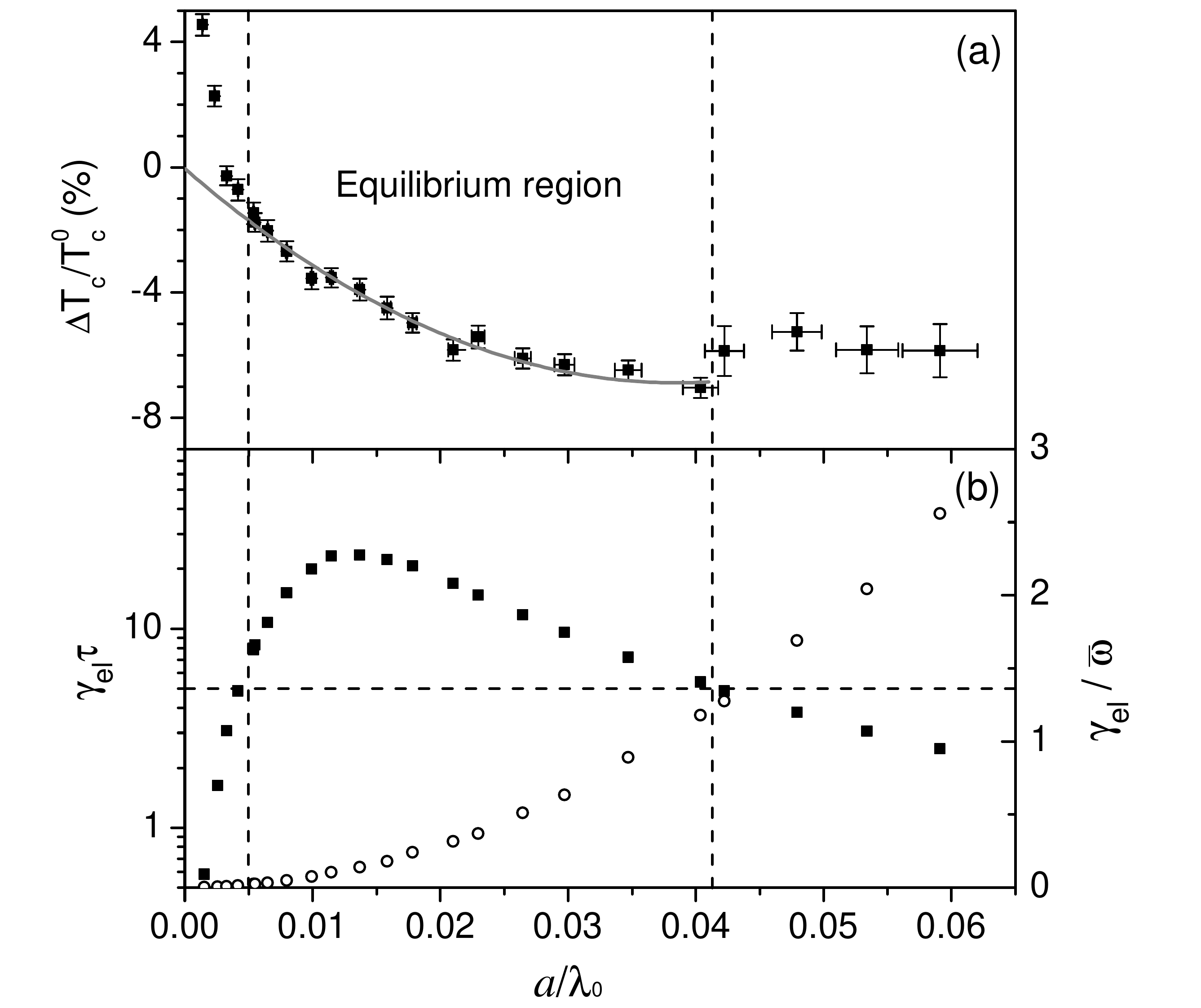}
\caption{Non-equilibrium effects.
(a) $\Delta T_c / \To$ is determined assuming equilibrium, as in Fig.~\ref{fig:Nc}. At both very low and very high $a$ the apparent $T_c$ deviates from the equilibrium curve. (b)
Equilibrium criteria (see text):
$\gamma_{\rm el} \tau$ (solid squares) is the number of elastic collisions per particle during $1\,\%$ atom-loss; $\gamma_{\rm el}/\bar{\omega} = 1$ (open circles) marks the onset of the hydrodynamic regime. (Figure adapted from \cite{Smith:2011}.)}
\label{fig:NonEq}
\end{figure}

An interesting question is then what happens if we violate these stringent equilibrium criteria.
In Fig.~\ref{fig:NonEq}(a) we show measurements extending beyond the equilibrium region shown in Fig.~\ref{fig:TcShift}, and in Fig.~\ref{fig:NonEq}(b) we plot the corresponding $\gamma_{\rm el} \tau$.
Individually, $\gamma_{\rm el}$ and $\tau$ vary vastly as a function of $a$ ($\gamma_{\rm el}$ increasing and $\tau$ decreasing) \cite{Smith:2011}, but the breakdown of equilibrium appears to occur at $\gamma_{\rm el} \tau \approx 5$ in both the low- and high-$a$ limit.

In the small-$a$ limit the apparent $T_c$ is significantly above the equilibrium curve.
We can qualitatively understand this effect within a simple picture. In this regime, losses are dominated by one-body processes which equally affect $N_0$ and $N'$. The net effect of equilibrating elastic collisions would therefore be to transfer atoms from the condensate to the thermal cloud.
However the dissipation rate is too high compared to $\gamma_{\rm el}$, and so
$N_0$ remains non-zero even after the total atom number drops below the equilibrium critical value $N_c$.

In the large-$a$ limit the initial breakdown of equilibrium again appears to result in condensates surviving above the equilibrium $T_c$. However the physics in this regime is much richer, with several potentially competing effects requiring further investigation. For example, three-body decay affects $N_0$ and $N'$ differently, the thermal component is far from saturation, and the thermal component of the gas also enters the hydrodynamic regime, $\gamma_{\rm el}/\bar{\omega} > 1$.

It is interesting that we observe ``superheated" BECs for both very weak and very strong interactions. Further
study of these effects should prove useful for improving the understanding of condensation in intrinsically out-of-equilibrium systems, such as polariton gases.

\end{document}